# Theoretical aspects of the study on the new bismuth chalcogenide based superconductors

Hidetomo Usui and Kazuhiko Kuroki
Department of Physics, Osaka University, 1-1 Machikaneyama-cho, Toyonaka, Osaka, 560-0043, Japan

**Abstract**
We review theoretical aspects of the studies on the recently discovered $BiCh_2$ (Ch=chalcogen) based superconductors. We first focus on the electronic band structure, the Fermi surface and their correlation with the lattice structure. We then discuss the phonon features and the issue of the lattice instability. Finally, we survey the phonon-mediated as well as the unconventional pairing mechanisms that have been proposed so far.

## 1. Introduction

Ever since the discovery of high $T_c$ superconductivity in the cuprates, a number of interesting superconductors possessing layered structure have been found. Among them are the organic superconductors [1,2], $MgB_2$ [3,4], $Sr_2RuO_4$ [5,6], $Na_xCoO_2$ [7,8], (Hf,Zr)NCl [9-11], and the iron-based superconductors [12,13]. Theoretically, the pairing mechanism and/or the pairing symmetry of these materials have been of great interest, and in order to clarify these issues, it is important to understand the electronic as well as the phonon features. Among these materials, the cuprates and the iron-based superconductors share a commonality in that they consist of conducting and insulating blocking layers, and many different materials can be synthesized by varying the blocking layer. In this regard, the newly discovered $BiCh_2$ (Ch: chalcogen) based superconductors form another group of layered materials with various kinds of blocking layers [14,15]. Mizuguchi *et al*. first discovered in 2012 $Bi_4O_4S_3$ with $T_c$ = 8.6 K [16]. $Bi_4O_4S_3$ has a layered structure consisting of $BiS_2$, $SO_4$ and BiO layers, and it was revealed from the first principles band calculation that the $BiS_2$ layer is the conducting layer. $BiS_2$ layers take a bi-pyramidal lattice structure reminiscent of those in some of the cuprate superconductors consisting of CuO pyramids [17]. After the discovery of $Bi_4O_4S_3$, $LaOBiS_2$ was immediately discovered [18]. This material has the same $BiS_2$ layers, but the blocking layer consists of LaO, which has the same structure as the in the iron based superconductor LaFeAsO [12]. Many materials have been found by substituting the lanthanide, e.g. $CeOBiS_2$ [19], $PrOBiS_2$ [20], $NdOBiS_2$ [21],

SrFBiS$_2$[22] and other materials [23-24]. Recently, bismuth selenide LaOBiSe$_2$ has also been found to show superconducting properties similar to the sulfides [25]. These materials are often referred to as the 1112 systems.

Superconductivity appears by doping electrons into the BiS$_2$ layers. In Bi$_4$O$_4$S$_3$, the electron doping is accomplished by controlling the amount of SO$_4$ molecules as (SO$_4$)$_{1-x}$. It has been found that Bi$_4$O$_4$S$_3$ ($x$ = 0.5) is metallic, while Bi$_6$O$_8$S$_5$ ($x$ = 0) is a band insulator [16,26]. In the 1112 materials $RE$OBi(S,Se)$_2$ (RE=rare earth), electron doping is realized by substituting fluorine for oxygen in the same way as in the iron based superconductors. Superconducting transition temperature $T_c$ is maximized at the fluorine doping level of $x$ = 0.5 and the maximum $T_c$ exceeds 10K for LaO$_{0.5}$F$_{0.5}$BiS$_2$ [27]. Chalcogen S can also be replaced by Se, which gives LaOBiSe$_2$ with $T_c$ = 2.6 K [25]. In this article, we give a review on the theoretical aspects of the studies on BiCh$_2$ based superconductors. Features of the crystal structures, band structures, phonon dispersions and candidates for the superconducting pairing mechanism will be discussed.

## 2. The crystal structure and band structure
### 2.1 Crystal structure

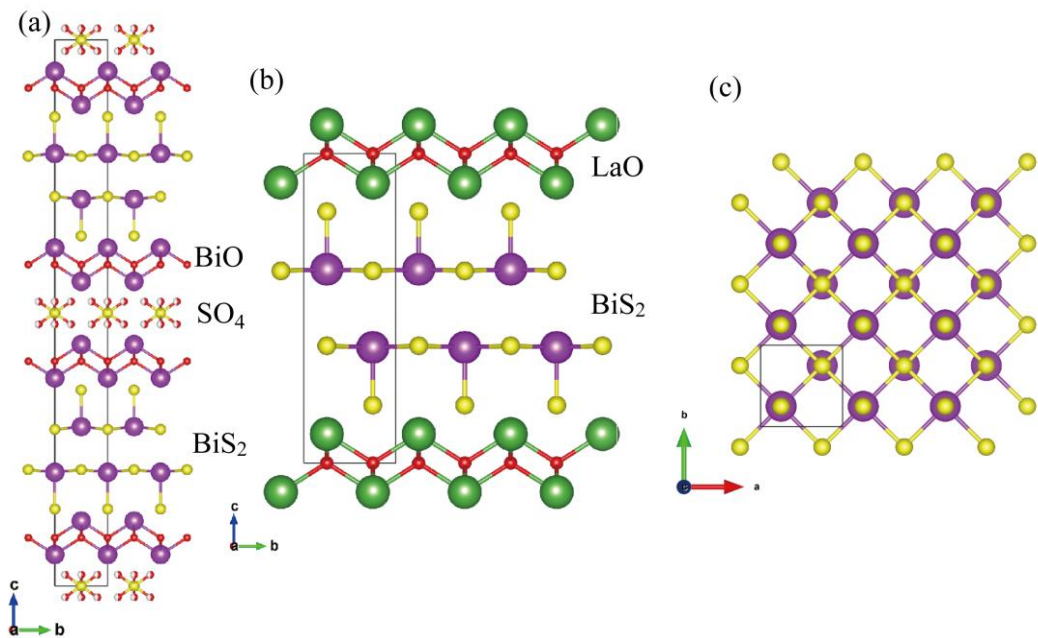

Fig. 1 The crystal structure of (a) Bi$_4$O$_4$S$_3$ and (b) LaOBiS$_2$, and (c) the BiS plane.

The crystal structure of BiCh$_2$ based superconductors consists of the Bi-Ch$_2$ bilayers and the blocking layers. The crystal structures Bi$_4$O$_4$S$_3$ and LaOBiS$_2$ are shown in Fig.1 (a) and (b), respectively. Here, we will focus only on the 1112 system. LaOBiS$_2$ has a tetragonal lattice structure belonging to the *P*4/*nmm* space group, with the lattice constants $a$ = 4.04 Å and $c$ = 13.8Å [28]. The BiS$_2$ bilayers have a pyramidal type structure consisting of the BiS plane and the apical S atoms, which is reminiscent of the bi-layer cuprates consisting of CuO pyramids [17]. The BiS plane exhibits a two-dimensional square lattice, as shown in Fig. 1(c). The BiS plane is not perfectly flat, and the atoms are buckled to some extent. The Bi-Ch-Bi bond angle, namely the strength of the buckling, is governed by the blocking layer [29] and also the amount of fluorine doping [30].

The lattice structure can also be determined by theoretical optimization[31-34]. These results indicate that the effect of the fluorine doping on the crystal structure is to increase the lattice constant $a$ and reduce the lattice constant $c$, and in total to reduce the unit cell volume. From the first principles calculations [31,32], it is shown that the BiS plane is flattened as the fluorine content increases from $x$ = 0 to 0.5. On the other hand, at $x$ = 1 the buckling of the BiS plane increases again. Therefore, not only the blocking layer, but also the amount of the electron doping controls the conduction layer. As will be discussed in section 2.3, theoretical studies have predicted dynamical or static lattice distortions [31-33]. Local distortion of Bi-S bond length has also been observed experimentally [35].

## 2.2 Electronic band structure

The band structure of BiCh$_2$ based superconductors has been obtained by first principles band calculations [16, 31-34, 36-41]. The band structures of Bi$_4$O$_4$S$_3$ and LaOBiS$_2$ calculated without spin orbit coupling are shown in Fig. 2 (a) and (b), respectively [16,36]. For Bi$_4$O$_4$S$_3$, the radii of the circles in Fig. 2(a) shows the strength of the Bi 6$p$ orbital character. The orbital character of the band structure of LaO$_{1-x}$F$_x$BiS$_2$ ($x$ = 0, 0.5 and 1) is shown in Fig. 3 [32]. The common feature of the band structure is that the conduction bands consist of the antibonding states of the Bi 6$p$ and S 3$p$ orbitals (mainly Bi 6$p$ orbitals). In LaOBiS$_2$, there are four conduction bands mainly composed of the Bi $p_x$ and $p_y$ orbitals, while in Bi$_4$O$_4$S$_3$, there are also other bands originating from other layers that contribute to the conductivity. We shall see later that the hopping integrals between the Bi 6 $p_x$ and $p_y$ orbitals are small, so that the conduction bands actually possesses one dimensional nature in spite of the apparently two dimensional feature of the crystal structure (see the modeling part in the latter part of this section). Due to the strong Bi-S hybridization, the valence bands have S $3p_x$, $3p_y$ character mixed with Bi $6p$, but we can see in Fig.3(a) that the valence bands just below the Fermi level around the Γ point do not have Bi $6p_{x/y}$ nor S $3p_{x/y}$ character for the mother compound ($x = 0$), which is because the origin of these bands is the blocking layers and the S $3p_z$ orbitals.

The conduction bands exhibit two-fold degeneracy along $k_x = \pi$ or $k_y = \pi$ (along X-M) due to the $P4/nmm$ symmetry of the 1112 system. In $Bi_4O_4S_3$, this two-fold degeneracy is lifted because of the $I4/mmm$ symmetry [16]. Even in the 1112 systems, the degeneracy is lifted along the X-Γ line, and this splitting reflects the magnitude of the hopping integral between the two layers within the Bi-$Ch_2$ bilayer structure (i.e., bilayer splitting). The splitting is small within the conduction bands because $p_x$ and $p_y$ orbitals of different layers barely overlap [36].

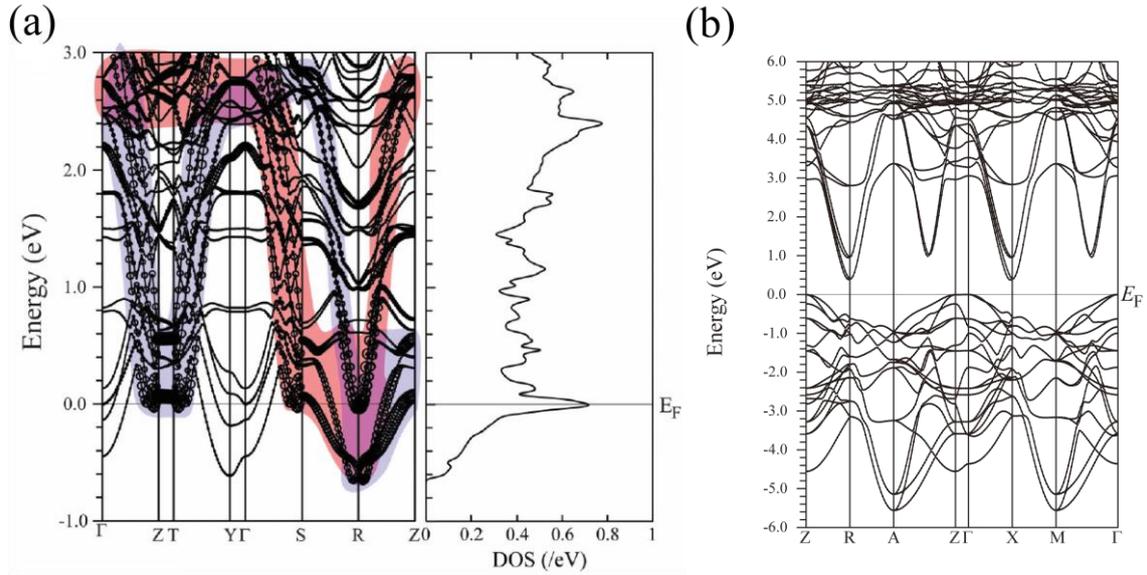

Fig. 2 The band structure of (a) $Bi_4O_4S_3$ [16] and (b) $LaOBiS_2$ [36].

As mentioned in subsection 2.1, the lattice constants and the internal coordinates vary with fluorine doping. The fluorine doping affects the band gap, band width and the details of the band structure. It can be seen from Fig.3 that the conduction bands around the Γ point shift downwards with respect to other bands as the fluorine is doped, and these bands eventually cross the Fermi energy (Fig. 3 (c)). Therefore, the number of Fermi surfaces is expected to increase when $x$ becomes close to 1, but such a large amount of doping is not realized in actual materials. Hence, within the realistic doping range, the rigid band picture is valid around the Fermi energy.

In Fig.4, the Fermi surface evolution of $LaOBiS_2$ is displayed, which was obtained within a rigid band approximation by varying the electron doping rate from $x = 0.2$ to 0.7 [37]. For all doping rates, the Fermi surface exhibits very weak three dimensionality, reflecting the layered crystal structure as well as the in-plane nature of the $p_x$, $p_y$ orbitals of the conduction bands. This is actually consistent with experimental observations on single crystals suggesting strong anisotropy of the conductivity[42]. The Fermi surfaces that appear around $(\pi,0)$ and $(0,\pi)$ are almost two-fold degenerate, where the splitting of the Fermi surface occurs due to the hoppings between the two BiS layers within the bilayer structure, i.e, the bilayer splitting. The density of states at the Fermi energy is maximized at around $x = 0.5$ because of the Lifshitz transition of the Fermi surface. For $x > 0.5$, the shape of the Fermi surface appears to be two dimensional as in the curates [43], but the orbital component of the Fermi surface possesses one dimensional character more like in $Sr_2RuO_4$ [44]. Note, however, that the band filling of $Sr_2RuO_4$ is around 1.33 per band on average, while that of $LaO_{0.5}F_{0.5}BiS_2$ is one-eighth filled per band.

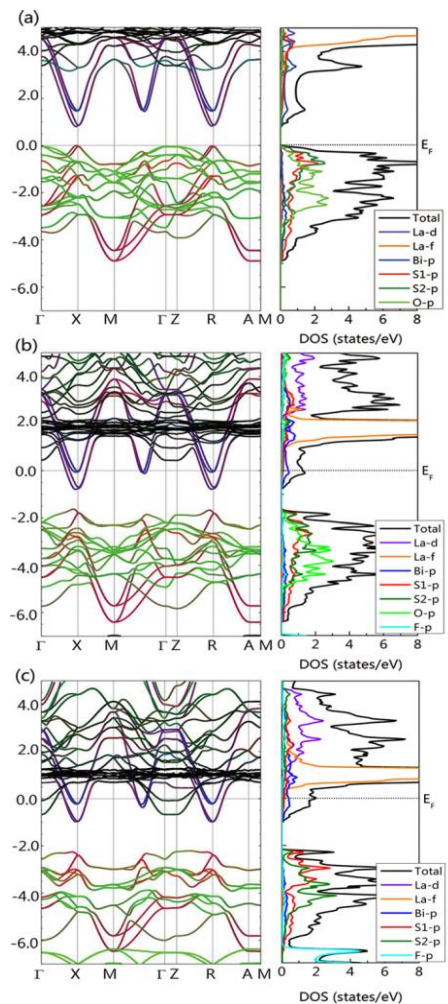

Fig. 3 The fluorine doping dependence of the band structure of LaO$_{1-x}$F$_x$BiS$_2$ [$x$ = (a) 0, (b) 0.5, and (c) 1] [32].

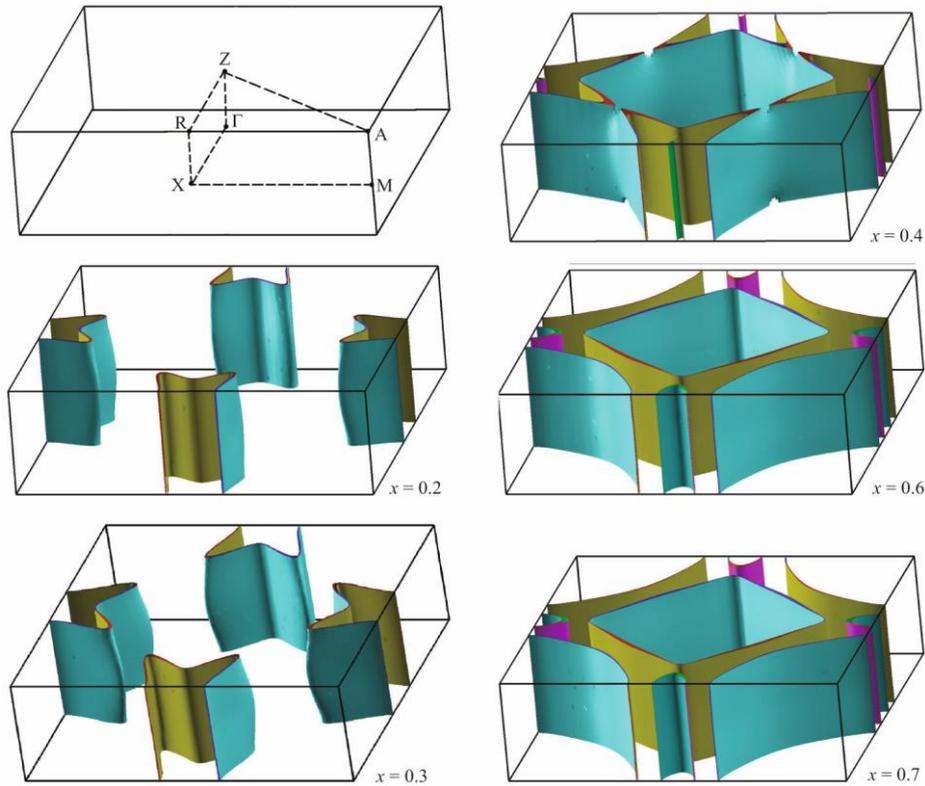

Fig. 4 The Fermi surface of LaO$_{1-x}$F$_x$BiS$_2$ without spin orbit coupling [37].

As far as the Fermi surface is concerned, the rigid band picture works well. However, the band gap itself is largely affected by fluorine doping, which can also be seen in Fig.3. For the non-doped material, the band gap is an indirect one between the band originating from Bi $p$ orbitals at the X point and that originating from the O $p$ orbitals at the Γ point. On the other hand, for $x$ = 0.5, the band gap becomes a direct one between the bands at the X point coming mainly from Bi $p$ and S $p$ orbitals within the BiS plane.

The indirect gap can be controlled by the onsite energy of O $p$ orbitals, which in turn is affected by the relative position between La and O in the blocking layer. The band structure assuming hypothetical lattice structure of LaOBiS$_2$ was calculated in ref. [39] as shown in Fig. 5, where the bond length between La and apical S atoms was varied. The indirect band gap increases upon increasing the bond length, namely, decreasing the onsite energy of the oxygen. The substitution of fluorine for oxygen also decreases the oxygen (fluorine) onsite energy. On the other hand, the direct band gap is controlled by the Bi-Ch intralayer hopping. When the Bi-Ch bond length becomes small by reducing the lattice constant *a*, the direct band gap decreases because the band width increases. The size of the band gap is also affected by the Bi-Ch-Bi bond angle, which controls the overlap of the Bi and Ch orbitals, and hence the hopping integral of the conduction bands.

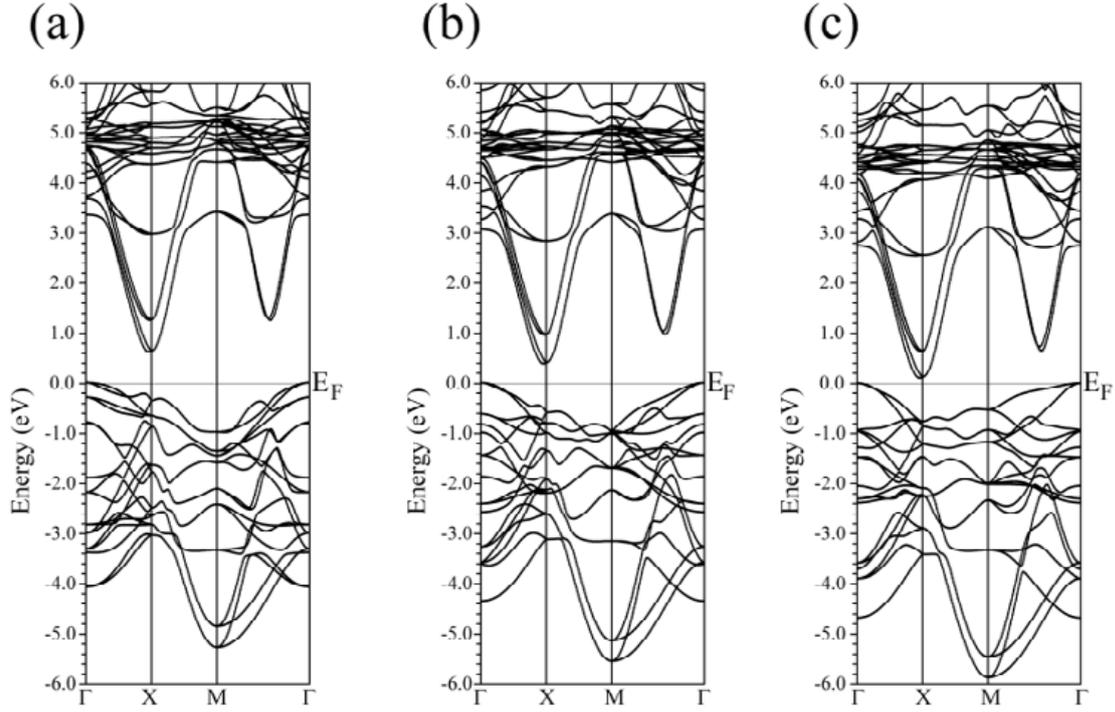

Fig. 5 The band structure as a function of $l_{\text{La-S}}$= (a) 4.11 Å, (b) 3.92 Å (original) and (c) 3.83 Å. [39]

So far we have seen the band structures calculated without spin-orbit coupling, but the Bi atom is generally known to have strong spin-orbit coupling effects. Fig.6 shows the comparison of the band structure and the Fermi surface calculated with and without taking into account the spin-orbit coupling [31]. With the spin-orbit coupling included, the band gap becomes small, and also the band splitting around the X point increases because the hopping integral between Bi-Ch layers within the bilayer structure increases. Hence, the bilayer splitting of the Fermi surface also increases with spin-orbit coupling.

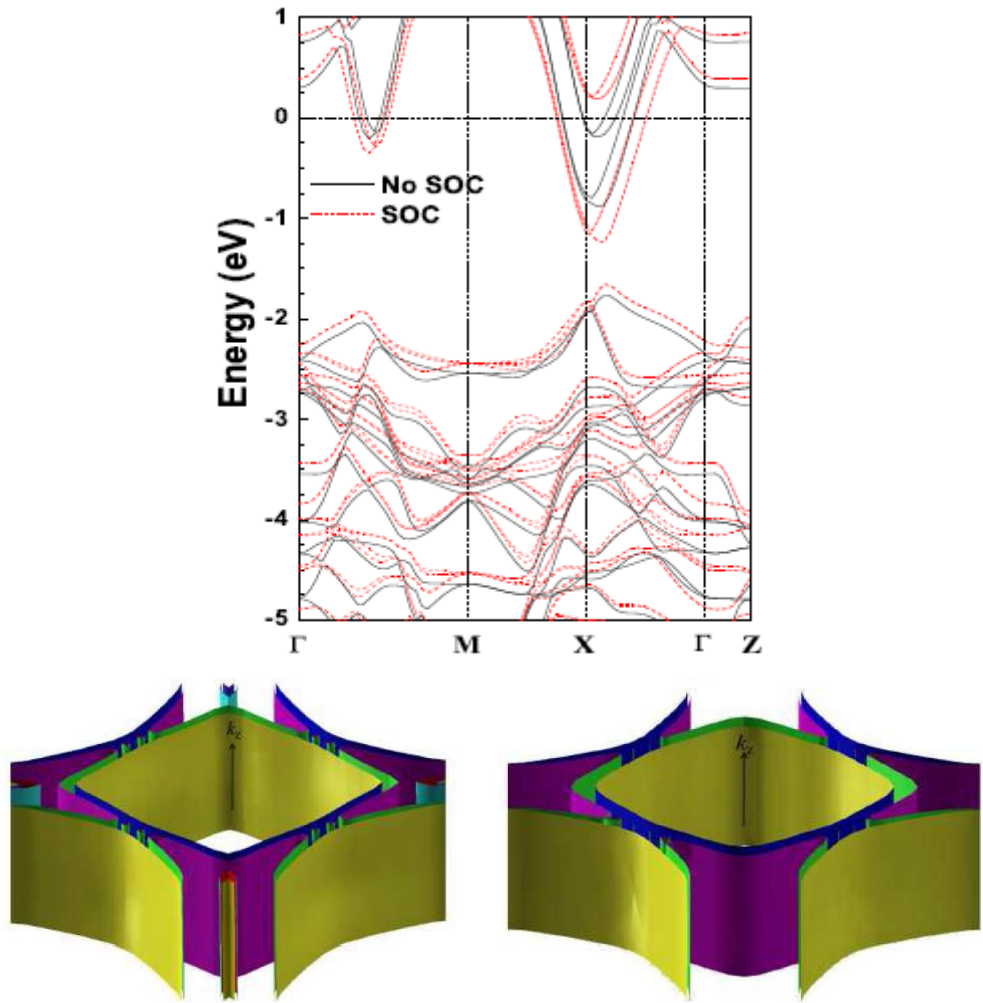

Fig. 6 Comparison of the band structure and the Fermi surface calculated with (right in the lower panel) and without spin-orbit coupling. [31].

We will now compare the calculation results with some experimental data. In Fig.7, we show the comparison between the first principles band calculation and the angle resolved photoemission (ARPES) result for $LaO_{1-x}F_xBiS_2$ with $x \sim 0.5$ [45]. At a glance, the agreement between experiment and theory looks good. The band width is nicely reproduced by the bare LDA calculation results, showing that the electron correlation effects are not so strong like in the 3$d$ systems such as the cuprates and the iron-based superconductors. Similar agreement between ARPES and the band calculation has also been found in ref.[46]. To be precise, however, there seems to exist some discrepancies between the theory and the experiment. For instance, in cuts (e) and (f) of Fig. 7(d), the theoretical expectation seems to be different from the experiment. Also, the ARPES Fermi surfaces of $NdO_{0.5}F_{0.5}BiS_2$ (Fig.8) [47] and $CeO_{0.5}F_{0.5}BiS_2$ [48] are found to be much smaller than the theoretical expectation of 50% fluorine doped compounds. In fact, the observed Fermi surface resembles that of the first principles band calculation assuming less than 20% doping. There may be some reason for the effective number of doped electrons to be much smaller than the nominal value.

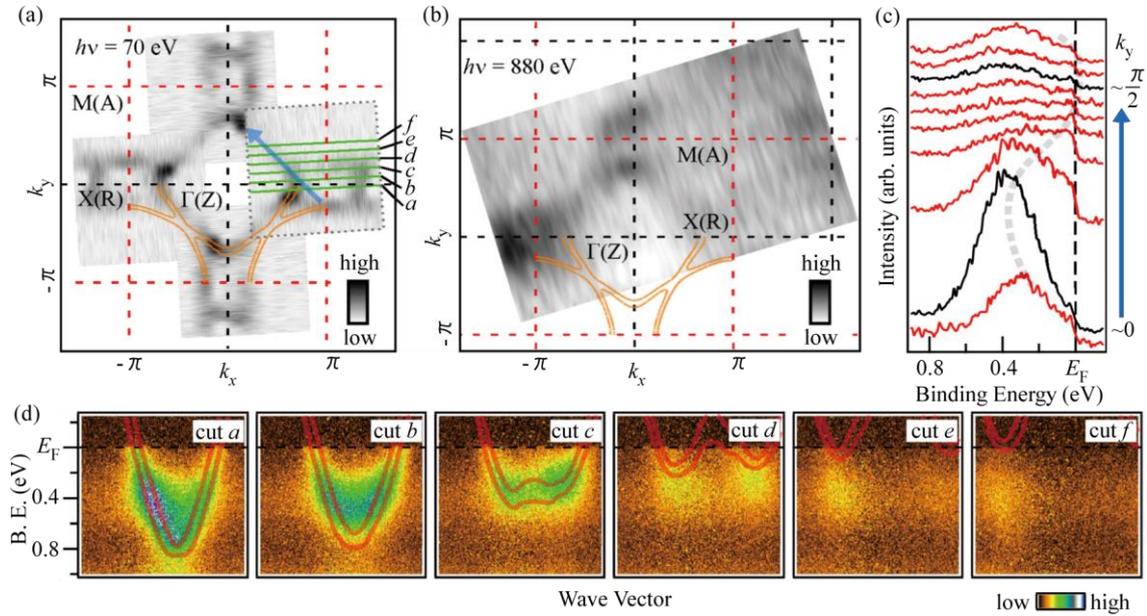

Fig. 7 (a) and (b) The observed Fermi surface with angle resolved photo emission spectroscopy, which is symmetrized according to the fourfold symmetry of the material. (d) Intensity plots as a function of the wave vector. The calculated bands and Fermi surface are superposed with solid lines [45].

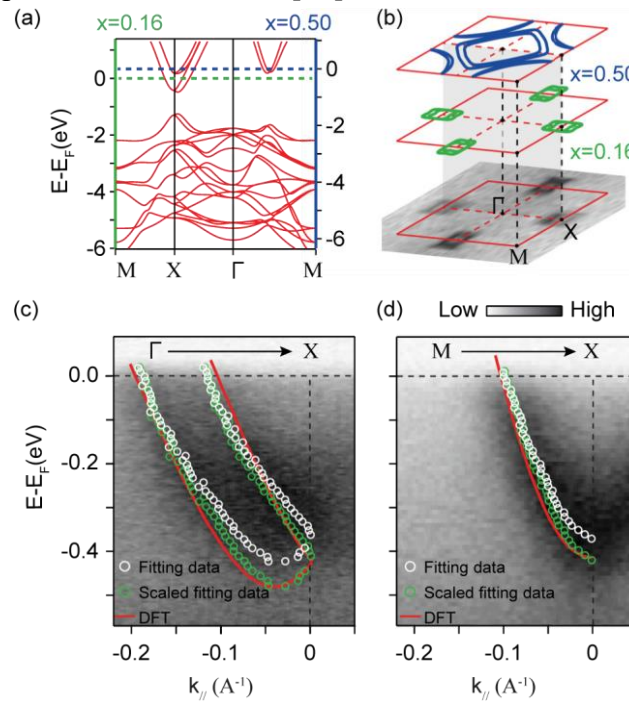

Fig. 8 (a) The band structure of NdO$_{1-x}$F$_x$BiS$_2$ calculated without spin-orbit coupling.

(b) the observed and the calculated Fermi surface for $x = 0.16$ and $0.5$. (c) and (d) The comparison of the band structure near $E_F$ between the photoemission data and the first principles band calculations along the Γ-X and M-X directions, respectively [47].

It is interesting to compare the electronic band features of BiCh$_2$ based superconductors to those of other bismuth based materials such as BaBiO$_3$ [49-51] or Bi$_2$Se$_3$ [52, 53]. (Ba,K)(Bi,Pb)O$_3$ becomes a superconductor with relatively high $T_c$ [51] and Bi$_2$Se$_3$ is famous as a topological insulator with a possibility of becoming a topological superconductor when doped with Cu [54, 55]. In BaBiO$_3$, the bands near the Fermi level originate from the Bi $6s$ orbital, so that the Fermi surface has a three dimensional shape [56]. In the case of Bi$_2$Se$_3$, the orbital character of the conduction bands is mainly $6p$ as in the BiCh$_2$ based compound, but the band structure exhibits strong three dimensionality [57].

We now move on to the modeling of the electronic structure [36]. As mentioned above, the band structure around the band gap is constructed from O, Bi, Ch $p$ orbitals, and the conduction bands mainly consist of the Bi $6p_x$ and $6p_y$ orbitals, which are strongly mixed with the S $3p_x$ and $3p_y$ orbitals. Due to the bilayer lattice structure, an eight-orbital model can be considered as suitable for describing the conduction states. The model can be further simplified by considering the fact that the hopping integral between the two layers within the bilayer structure is small due to the strong in-plane character of the $p_{x/y}$ orbitals. Then, adopting a single layer approximation, we can construct a four-orbital model originating from Bi 6 $p_{x/y}$ and S 3 $p_{x/y}$ orbitals as shown in Fig. 9(a). Finally, considering only the conduction bands, we can construct a two-orbital model as shown in Fig.9(b). This two-orbital model has been widely used for theoretical studies of unconventional pairing by adding the electron-electron interaction and/or the spin-orbit coupling terms [58-67].

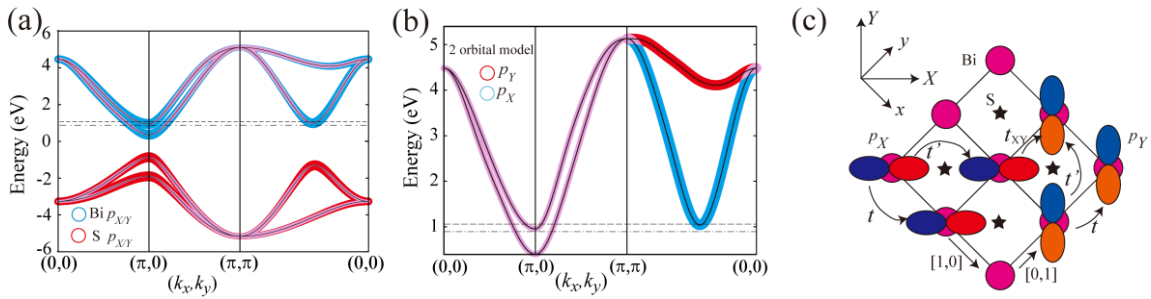

Fig. 9 The band structure of (a) the effective two-orbital, (b) the four-orbital models, and (c) the tightbinding model of the BiS plane [36].

In order to discuss the magnitude of the hopping integrals, we adopt the $p_X$ and $p_Y$ orbitals, where the *X-Y* axes are rotated by 45° (see Fig. 9(c)) [36]. This is because the largest hopping integral is along the Bi-S bonding direction, namely, the *X* and *Y* axes. The two orbital model mainly consists of the nearest and the next nearest neighbor intraorbital hoppings $t$ and $t'$, and the nearest neighbor interorbital hopping $t_{XY}$ (Fig. 9(c)). The values are $t$=-0.167, $t'$= 0.88 and $t_{XY}$ = 0.107eV for LaOBiS$_2$ [36]. The mixture between the $p_X$ and $p_Y$ orbitals ($t_{XY}$) is relatively small, and this mixing controls the two dimensionality of the system. The Fermi surface at a fluorine doping ratio of $x$ = 0.2 in Fig. 4 exhibits a two dimensional feature due to this $p_X$-$p_Y$ mixing.

Although this two-orbital model almost perfectly reproduces the band structure and the Fermi surface of the first principles calculation, the hopping integral between the Bi-Ch layers should be included if the effect of the spin-orbit coupling on the band structure is taken into account. The spin-orbit coupling slightly increases the bilayer splitting of the Fermi surface [31]. Then, the four-orbital (conduction bands only) or the eight-orbital (conduction and valence bands) models are more suitable for describing the effect of the mixture of the orbitals between the layers.

## 2.3 Phonon features

The phonon dispersion has been calculated for LaOBiS$_2$ [31-33] and LaOBiSe$_2$ [34]. In the left panel of Fig.10, we show the phonon dispersion calculated in ref.[33] for the *P*4/*nmm* lattice structure of LaO$_{1-x}$F$_x$BiS$_2$ at $x$ = 0 and 0.5 using a 4 × 4 × 1 supercell. Negative energy, hence unstable, phonon modes are seen for both doping levels.

The phonon modes at Γ and M points can be divided into 6$B_2$, 10$E$ and 4$A_1$ at Γ point and 4$B_1$, 4$A_1$, 5$B_2$ 5$A_2$ and 6$E$ at M point [32]. At $x$ = 0, the orthorhombic *P*21*mn* symmetry is the most stable structure due to the *E* mode at Γ, where the S atoms move towards Bi atoms along the *a* or *b* axis [33]. The total energy of this structure is ~1 meV smaller than that of the original lattice structure as shown in the right panel of Fig.10. The potential energy is therefore very shallow, so that the tetragonal lattice structure should be observed because of the quantum fluctuations, in agreement with the actual experiments.

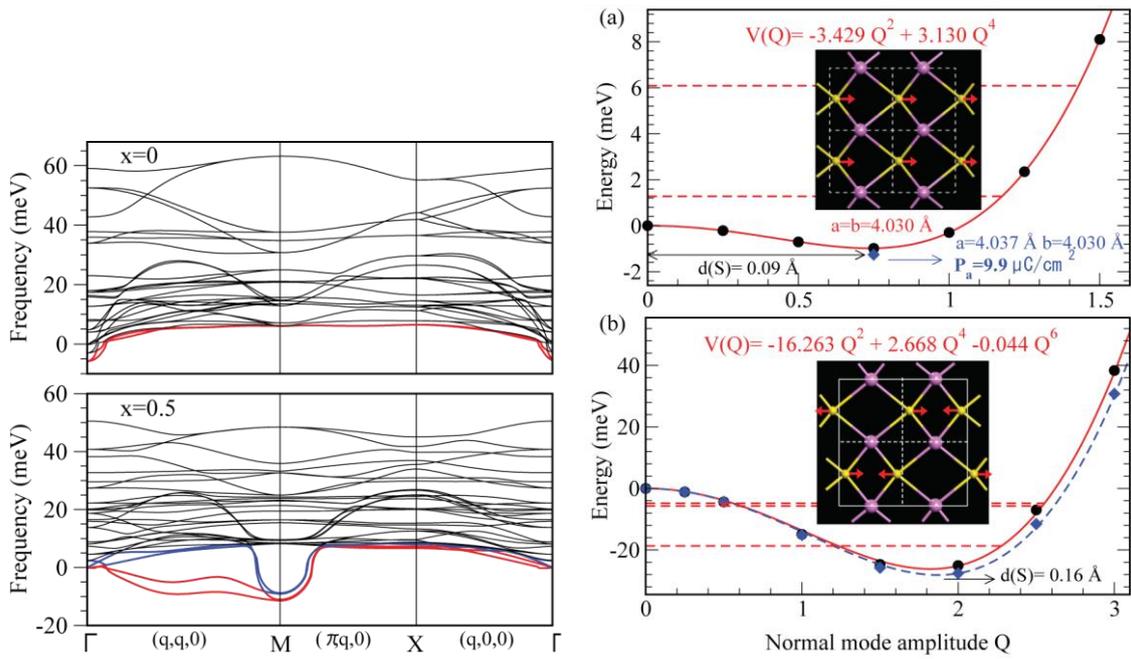

Fig. 10  Left panel : the phonon dispersion of LaO$_{1-x}$F$_x$BiS$_2$ at $x = 0$ and 0.5. Right panel : energy variation when the system is distorted by the negative energy phonons. (a) at Γ for x=0, (b) at M for x=0.5. [33].

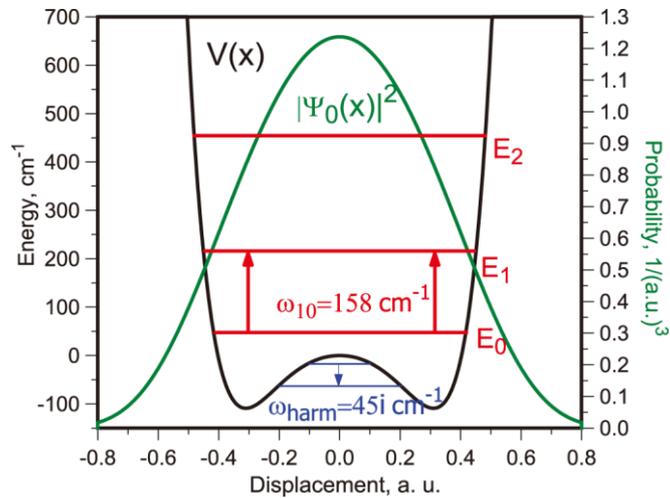

Fig. 11 Calculated double well potential for the frozen phonon mode as a function of the displacement from the symmetric point in the tetragonal structure, along with the probability of the ground state atomic wave function [31].

At $x = 0.5$, the unstable mode is present at the M point, namely the wave vector $(\pi,\pi)$. The potential energy against the distortion corresponding to the lowest phonon mode has a much deeper local minimum compared to the distortion for $x = 0$. In ref.[33], energy levels that are bound to this local minimum were obtained (Fig.10 right panel), while in ref. [31], the obtained energy levels were unbound (Figs. 11), so that the system should appear as tetragonal. In any case, the system is close to the border between dynamically or statically distorted lattice structure.

The origin of the unstable modes for $x = 0$ and 0.5 is different. In ref. [33], it has been suggested that the ferroelectric unstable mode at $x = 0$ occurs due to the mismatch of the optimum lattice constants between LaO and $BiS_2$ layers. This was confirmed by recalculating the phonon dispersion with a smaller lattice constant $a$ ($a$ = 3.8Å), for which the soft phonon mode was not obtained. On the other hand, at $x = 0.5$, it was shown that taking a smaller lattice constant does not stabilize the negative energy phonons.

Actually, the unstable mode at $x = 0.5$ is more related to the Fermi surface nesting. In ref.[33], it was shown that this unstable phonon mode actually spreads along the Γ-M line. The relation between this and the Fermi surface nesting can be clearly seen in the two orbital model described in subsection 2.2. The largest eigenvalue of the irreducible susceptibility of the two-orbital model at $x$=0.5 is shown in Fig. 12(a) [36]. The diagonal structures that go through (0,0) or (π,π) are due to the Fermi surface nesting shown by the arrows in Fig. 12 (b). Along the ($q,q,0$) line, the irreducible susceptibility is enhanced because of the one dimensional nature of the $p_X$ and $p_Y$ orbitals. Hence, the unstable phonon dispersion along Γ-M reflects the unstable nature of the electronic system along this line, which implies a strong electron phonon coupling.

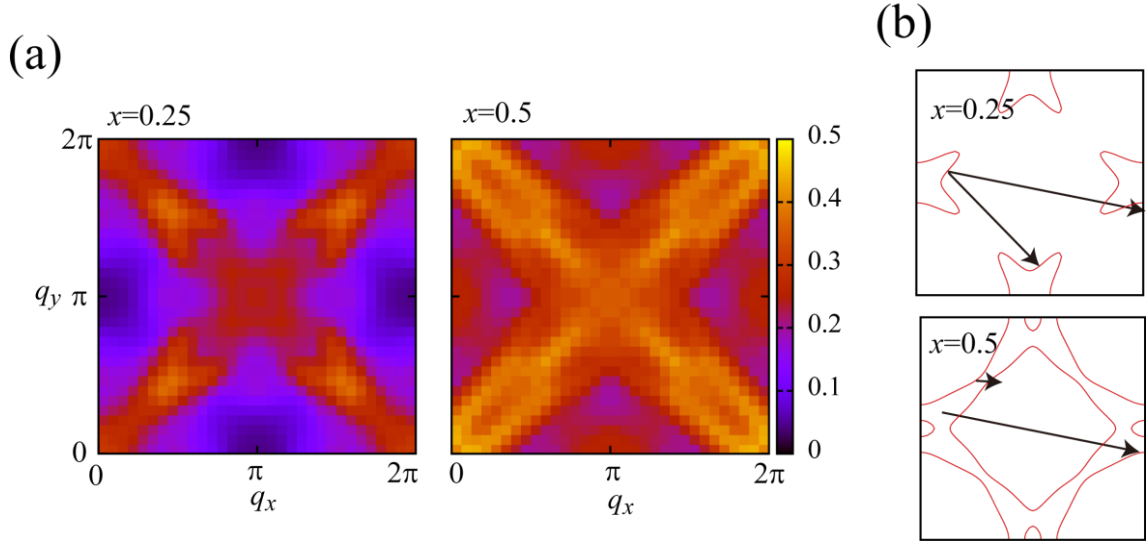

Fig. 12 (a) The irreducible susceptibility with the two orbital model of $LaO_{1-x}F_xBiS_2$ [36], and (b) the Fermi surface and the nesting vector.

The phonon features of LaO$_{0.5}$F$_{0.5}$BiSe$_2$ are basically similar to that of the sulfur counterpart [34]. However, the potential energy minimum of the ($\pi,\pi$) distortion is much shallower. In the case of the selenide, it is unlikely that the distortion is statically stabilized.

## 3. Superconductivity

The pairing mechanism of the BiCh$_2$ based superconductors is still under debate. In this section, we will discuss some of the pairing mechanism candidates.

### 3.1 Expriemental results

First, we briefly summarize some of the experimental results. It has experimentally been found that $T_c$ of LaO$_{1-x}$F$_x$BiS$_2$ is maximized at $x = 0.5$ [27]. The maximum $T_c$ at around $x = 0.5$ has also been observed in other 1112 systems. $T_c$ is also dependent on the lattice parameters within the 1112 systems. This may be expected because different rare earth blocking layer gives slight difference in the lattice constants as well as the buckling within the BiCh planes. Quite recently, it has been shown that $T_c$ is systematically correlated with the chemical pressure, which has been quantitatively defined as the in-plane distance between Bi and Ch atoms normalized by the summation of their ionic radii [68].

As for the pairing symmetry, experimental results have suggested the fully gaped state, which is consistent with *s*-wave pairing [69-72]. For example, the penetration depth measurements indicate that the superfluid density can be fit by theoretical calculations assuming multiple *s*-wave gaps [69-71]. Strong coupling superconductivity has been indicated in a number of studies, but the value of $2\Delta/k_B T_c$ varies among various measurements ; for instance, very large values such as 7.2 for Bi$_4$O$_4$S$_3$ (penetration depth measurement) [69] and 16.8 for Nd(O,F)BiS$_2$ (scanning tunneling spectroscopy) [73] was reported, while the above mentioned penetration depth measurement for Nd(O,F)BiS$_2$ observed  $2\Delta/k_B T_c$ ~4 [71].

### 3.2 Phonon mediated pairing

The possibility of phonon-mediated pairing has been investigated from the early stage of the theoretical studies [31-34]. Based on the first principles calculation results, $T_c$ can be estimated using the Allen-Dynes formula [74, 75], $T_\mathrm{c} = \frac{\omega_\mathrm{ln}}{1.2}\exp\left[-\frac{1.04(1+\lambda)}{\lambda-\mu^*(1+0.62)\lambda}\right]$, where $\lambda$ is the electron-phonon coupling, $\mu^*$ is the screened Coulomb interaction parameter, and $\omega_\mathrm{ln}$ is the logarithmically averaged phonon frequency. The electron phonon coupling λ can be calculated as $\lambda = 2\int_0^\omega \frac{\alpha^2 F(E)}{E} dE$, where $\alpha^2 F$ is the Eliashberg function. In all of the calculations, the electron-phonon coupling is estimated to be large. For the tetragonal lattice structure, λ = 0.85 was obtained for $LaO_{0.5}F_{0.5}BiS_2$ [31] and λ = 0.51 [34] for $LaO_{0.5}F_{0.5}BiSe_2$. The difference comes from the contribution of the anharmonic, unstable phonons, for which the sulfide has a deeper potential minimum and a larger distortion. These calculations give $T_c$ = 9~11K for La(O,F)BiS$_2$ [31-33] and $T_c$ ~ 2.4 K for La(O,F)BiSe$_2$ [34], assuming $\mu^*$ = 0.1.

In ref.[33], the electron-phonon coupling was calculated for two distorted lattice structures of $LaO_{0.5}F_{0.5}BiS_2$, namely, λ = 0.83 for the (π,π) optimized phase and λ = 0.6 for the CDW phase [33]. The phonon density of states, the Eliashberg function $\alpha^2 F$ and the electron-phonon coupling constant λ calculated in ref. [33] are shown in Fig. 13.

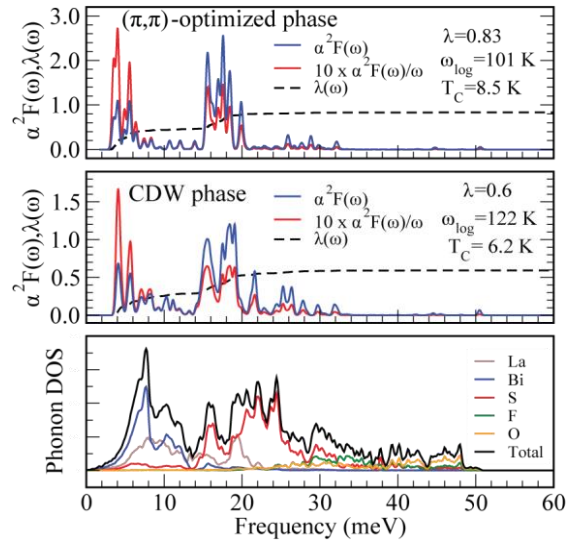

Fig. 13 The phonon density of states, the Eliashberg function $\alpha^2 F$ along with the electron-phonon coupling constant λ (dashed lines) calculated for the two distorted lattice structures of $LaO_{0.5}F_{0.5}BiS_2$ [33].

In both phases, the Eliashberg function $\alpha^2 F$ exhibits similar features. The electron phonon spectral function is large in the low frequency regime around several meV, and also in the intermediate frequency regime around 20 meV. The phonon density of states shows that the low frequency modes originate from Bi (and La) vibrations, while the intermediate frequency modes around 20 meV are from S vibrations. According to ref. [32], one of the low frequency phonon modes is the $A_1$ mode, which corresponds to the vertical vibrations with the upper and lower BiS$_2$ layers moving in the opposite direction, and two of the intermediate frequency modes are the $E$ and $B_2$ modes. In the $E$ mode, LaO and BiS$_2$ layers move toward the $x$ or $y$ axis. In the $B_2$ mode, La, Bi and in-plane S atoms move along the same vertical direction, while the apical S in the BiS$_2$ plane and O atoms move in the opposite directions of La, Bi and in-plane S atoms.

The large electron-phonon couplings obtained for BiCh$_2$ based superconductors are comparable to that of MgB$_2$ [3,4], which has the highest $T_c$ among conventional superconductors at ambient pressure. Within the phonon-mediated pairing scenario, the reason why $T_c$ of the BiCh$_2$ based superconductors is much lower than that of MgB$_2$ is because of the small logarithmic phonon frequency average. The logarithmic frequency average is 100~260K for LaOBiS$_2$ [31-33], while estimations for MgB$_2$ have given ω = 540K [76] and 700K [77].

These calculated $T_c$ values appear to be in good agreement with the experimental observations (few K ~ 10K), and the phonon-mediated $s$-wave state is consistent with the experiments mentioned in the beginning of this section. Nonetheless, there are some controversies regarding the strong electron-phonon coupling scenario. For instance, a Raman scattering experiment estimates the electron-phonon coupling to be large, consistent with the theories [72], while another Raman experiment, with different kind of analysis, gives a smaller estimation, suggesting unconventional pairing [78]. Also, applying chemical [68,79,80] or physical pressure [81-83] gives rise to enhancement of $T_c$, but this may be in contradiction with a simple BCS expectation because the application of pressure should reduce the density of states. Moreover, a neutron scattering study of La(O,F)BiS$_2$ has shown that the phonon modes are not affected by the normal to superconducting transition, in contradiction to the theoretical expectation of a strong electron-phonon coupling superconductor [84]. One should also note that the theoretical estimations of strong electron-phonon coupling have been obtained for the fluorine content of $x$=0.5, where the Fermi surface nesting is good and the Fermi level sits at the peak of the electronic density of states, but the ARPES experiments for NdO$_{0.5}$F$_{0.5}$BiS$_2$ (Fig.8) [47] and CeO$_{0.5}$F$_{0.5}$BiS$_2$ [48] suggest that the effective electron doping rate is much smaller than that expected for $x$ = 0.5, as mentioned previously. In fact, the Sommerfeld coefficient of the specific heat γ is estimated to be smaller (less than 2.5mJ/mol K$^2$) than the theoretical estimation for LaO$_{0.5}$F$_{0.5}$BiS$_2$ (3mJ/molK$^2$) [15] suggesting both smaller carrier concentration and weak electron-phonon coupling.

### 3.3 Unconventional pairing mechanisms

Unconventional pairing mechanisms have also been proposed for the BiCh$_2$ based superconductors. Several studies have examined the spin-fluctuation-mediated pairing scenario [36,61,64,67]. If one assumes a short range (e.g., on-site only) interaction, strong spin fluctuation may arise due to the Fermi surface nesting. Applying the random phase approximation to the two-orbital model of Bi $6p_x$, $p_y$ orbitals (in which the spin-orbit coupling is omitted), the competition among extended $s$, $d_{xy}$, $d_{x2-y2}$, and $g$-wave pairings have been discussed [36,61,64,67]. Fig.14 shows the calculation result of ref. [64]. Here, the eigenvalue $\lambda$ of the Eliashberg equation is a qualitative measure of the superconducting transition temperature (note that $\lambda$ here is not the electron-phonon coupling constant). In the small doping regime around $x$ = 0.14, the $g$-wave ($A_{2g}$) symmetry is dominant. However, in the strong Hund's coupling regime, $d$-wave, $g$-wave and $s$-wave ($A_{1g}$) are in close competition. In the large doping regime, the results show that the $g$-wave pairing competes with both the $s$-wave ($A_{1g}$) and $d$-wave ($B_{2g}$) pairing, and no pairing symmetry emerges dominantly as shown in Fig.15.

In the spin-fluctuation-mediated pairing, the sign of the superconducting gap should change across the nesting vector along the $(q,q,0)$ line. In fact, the superconducting gap obtained within this formalism all have nodal structure on the Fermi surface. Hence, the experimental observation of a fully open superconducting gap cannot be explained within these scenarios. On the other hand, magnetic-interaction-mediated pairing has been examined in a model where the nearest and the next nearest neighbor antiferromagnetic interactions ($J_1$, $J_2$) between localized spins are considered. There, a superconducting state with fully gapped extended $s$-wave pairing was obtained [60]. Taking into account the spin-orbit coupling does not basically affect these results [60].

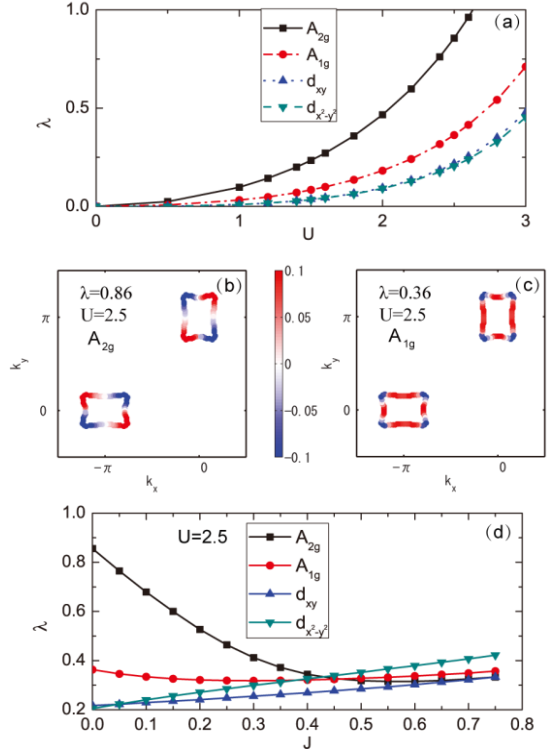

Fig. 14 The eigenvalue $\lambda$ of the Eliashberg equation for the $A_{2g}$, $A_{1g}$, $d_{xy}$ and $d_{x2-y2}$ pairing symmetries at $x = 0.14$ as a function of (a) on-site Coulomb interaction U and (d) Hund coupling $J$. The superconducting gap on the Fermi surface for (b) $A_{2g}$ and (c) $A_{1g}$ pairing symmetries [64].

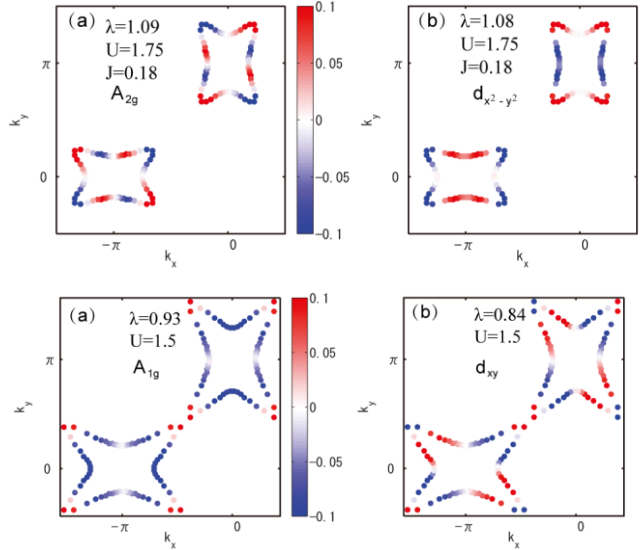

Fig. 15 Upper panels : the superconducting gap on the Fermi surface at $x = 0.25$ for (a) $A_{2g}$ and (b) $d_{x2-y2}$ pairing symmetries. Lower panels: the superconducting gap on the Fermi surface at $x = 0.5$ for (a) $A_{1g}$ and (b) $d_{xy}$ pairing symmetries [64].

The effect of the spin-orbit coupling on superconductivity has also been studied in refs.[63,65,66]. In ref.[63], the functional renormalization group method was applied to the two-orbital model in the presence of the spin-orbit coupling, There $d^*_{x2-y2}$-wave state was found to have the leading pairing instability, where "*" stands for the simultaneous rotation operation of both the spin and the lattice. This pairing respects the time reversal symmetry, and the triplet and singlet component are mixed, where the former dominates. This can be categorized as a time-reversal-invariant weak topological superconductor [63]. When the band filling is below the Lifshitz transition point, i.e., when two disconnected Fermi surfaces exist around $(\pi,0)/(0,\pi)$, the sign of the superconducting gap changes sign between the two Fermi surfaces, so that the gap is fully open as shown in Fig. 16. The calculated superfluid density based on this state is in good agreement with the experimental results (Fig. 17) [66]. An interesting prediction of this scenario is that the superconducting gap closes on the edge states because the system belongs to the category of weak topological superconductors.

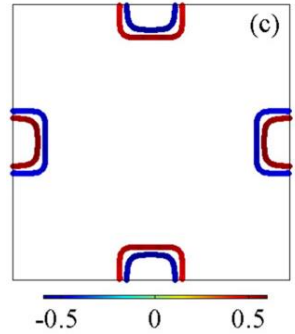

Fig. 16 The gap function of NdOBiS$_2$ obtained within the functional renormalization group method [66].

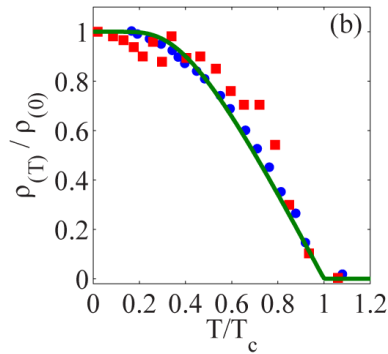

Fig. 17 The calculated superfluid density along with the experimental results for

Bi$_4$O$_4$S$_3$ and LaOBiS$_2$ [66].

The existing theoretical studies basically use the two-orbital model, which consider only one BiS plane. Therefore, the effect of the hoppings between the layers is not taken into account. The interlayer hopping barely affects the Fermi surface, but the bilayer lattice structure itself may affect superconductivity in a manner where the crystal symmetry plays some important role.

Before closing this subsection, we will briefly mention some of the experimental observations that may be related to the possibility of (other types of) unconventional pairing mechanisms. In ref.[73], the large $2\Delta/k_BT_c$ was interpreted as an indication of strong pairing interaction and the presence of superconducting fluctuation well above $T_c$. Suppressing the superconductivity by the magnetic field revealed unexpected semiconducting behavior in the normal state. These observations were taken as a possible indication that the pairing is related to charge density wave or valence fluctuations. Speaking of charge fluctuations, local structural distortions that accompany long and short Bi-S bond distances (either ferro or antiferro distortive) have been observed in an neutron scattering experiment for polycrystals of La(O,F)BiS$_2$ [35]. This local distortive fluctuation was linked to charge disproportionation of the Bi ions, which in turn may be related to the pairing mechanism. Also, scanning tunneling spectroscopy measurement on NdO$_{1-x}$F$_x$ BiS$_2$ has revealed a "checkerboard stripe" charge ordering on the surface [85]. This ordering forms one dimensional stripe structure along the Bi-Bi direction.

For SrFBiS$_2$ [22], an insulator to superconducting transition has been observed. Namely, the superconducting phase abruptly appears as soon as the metallicity sets in with electron doping by partially substituting Sr with La, and within the superconducting phase, $T_c$ interestingly increases with decreasing the electron doping content [86]. Similar behavior has been observed in layered nitride superconductors *M*NCl (*M*=Hf,Zr) [11], where unconventional pairing has been discussed. In *M*NCl, the conduction bands originate from 4*d* or 5*d* orbitals, so that the Fermi surface and its orbital character are different from those of the BiCh$_2$ based superconductors. However, the above experimental observation might suggest a commonality possibly connected to unconventional pairing. Other experimental observations have also suggested the possibility of unconventional pairing mechanism, e.g. structural instability [35] and structural phase transitions under pressure [87]

**3.4 Proposals for probing the pairing state**

Various experimental probes to determine the pairing state have been proposed, among which are the impurity effect [59], spin excitation [58] and nuclear magnetic resonance [66]. In unconventional superconductors with sign reversing order parameters, $T_c$ is reduced upon increasing the amount of non-magnetic impurities. In ref. [59], it was shown that $T_c$ is strongly suppressed against the impurity content for the *p* and *d*-wave pairings, while $T_c$ hardly decreases in the s-wave state. In ref.[58], the imaginary part of the spin susceptibility of various pairing states was calculated as a probe to determine the pairing state. It was shown that the *s*-wave shows no spin excitation, while spin excitations appear at the incommensurate momentum (0.7π, 0.7π) for d-wave, around (0,0) for p-wave. Ref. [66] studied NMR as a probe to detect the $d^*_{x2-y2}$-wave state. Although both *s*-wave and $d^*_{x2-y2}$-wave pairings are fully gapped states (below the Lifshitz point for $d^*_{x2-y2}$-wave), the magnitude of the coherence peak just below $T_c$ in the spin-lattice relaxation rate $1/T_1$ is significantly different between the two states due to the sign reversal of the order parameter in the latter. Namely, a large coherence peak appears in the s-wave, while it is suppressed in the $d^*_{x2-y2}$-wave. Since the $d^*_{x2-y2}$-wave state has predominant a spin-triplet pairing component, one might expect that the Knight shift can also be used as a probe to detect such a state. The calculation in ref. [66] however shows that the difference between *s*-wave and $d^*_{x2-y2}$-wave is insignificant.

## 4. Conclusion

In this review article, we have surveyed the features of the BiCh$_2$ based superconductors from a theoretical point of view. The BiCh$_2$ based superconductors are a group of layered superconductors constructed from the pyramidal BiCh$_2$ bilayers, and blocking layers having tetragonal structure. In 1112 systems, the electrons can be doped into the BiCh plane with fluorine substitution for oxygen. The conduction bands consist of the *p* orbitals within the BiCh plane. The conduction bands and the Fermi surface therefore have one dimensional nature with small hoppings along the vertical directions. The irreducible susceptibility is enhanced along the (*q,q*,0) line due to the Fermi surface nesting and its one dimensional nature. The calculated phonon dispersion in the tetragonal lattice structure also shows the soft phonon modes along the (*q,q*,0) line. This indicates the presence of strong electron-phonon coupling in these materials. The first principles electronic band calculation results are overall in good agreement with the ARPES experiment, suggesting absence of strong electron correlation as in the 3d-based layered superconductors. On the other hand, there do exist some discrepancies between theory and experiment, whose origin remains to be clarified.

The pairing mechanism has been studied in the context of both conventional and unconventional scenarios. $T_c$ values comparable to the experimental observations have been obtained within the phonon-mediated pairing owing to the strong electron-phonon coupling. As for the unconventional pairing, the weak coupling spin-fluctuation calculations omitting the spin-orbit coupling predicts nodal gap functions, while a fully gapped *s*-wave state is obtained for the $J_1$-$J_2$ local spin model. In a model with spin-orbit coupling, a functional renormalization group study predicts a $d^*_{x2-y2}$-wave pairing, which belongs to the category of the weak topological superconductivity. In this pairing, the triplet and the singlet pairing components are mixed, and the order parameter changes its sign between the disconnected Fermi surfaces. Experimental results also suggest the possibility of charge-fluctuation-mediated superconductivity.

As surveyed in this review, the $BiCh_2$ based superconductors have interesting features, such as the one dimensional nature of the electronic bands and the Fermi surface nesting, strong electron-phonon coupling, and the strong spin-orbit coupling, which may be entangled in a complex manner. The issue of the pairing mechanism and the pairing symmetry should be closely related to these features. In order to have better understanding on these issues, further theoretical as well as experimental studies are necessary.

**Acknowledgement**

We are grateful to Katsuhiro Suzuki, Yoshikazu Mizuguchi, Yoshihiko Takano, Hiroshi Fujihisa, Hisashi Kotegawa, Takashi Yokoya, Kensei Terashima, Ryotaro Arita, Akihiro Kokubo, Hiroaki Tsuchiya, Yuki Fuseya, George Martins, Hai-Hu Wen, Dong-Lei Feng, Naurang Saini, Yuichi Ota, Kozo Okazaki, Takuya Sugimoto, and Takashi Mizokawa for fruitful discussions. This study has been partially supported by Grants-in-Aid for Scientific Research No. 26610101 and 26247067 from the Japan Society for the Promotion of Science.